\documentclass{amsart}
\usepackage{amssymb}
\usepackage{amsfonts}
\usepackage{amsmath}
\usepackage[legalpaper,bookmarks=true,colorlinks=true,linkcolor=blue,citecolor=blue]{hyperref}
\usepackage{graphicx}%
\usepackage{fancyhdr}
\usepackage{color}
\usepackage[mathlines]{lineno}
\usepackage{lscape}
\usepackage{epsfig}
\usepackage{natbib}
\usepackage{geometry}
\usepackage{tgbonum}
\usepackage[]{listings}

\fontfamily{qcr}\selectfont

\newtheorem{theorem}{Theorem}
\theoremstyle{plain}

\newtheorem{lemma}{Lemma}

\numberwithin{equation}{section}

\begin{document}
\title[Functional Empirical process]{How to use the functional empirical process for
deriving asymptotic laws for functions of the sample}
\author{Gane Samb LO}
\address{LSTA, Universit\'e Pierre et Marie Curie, France and LERSTAD,
Universit\'e Gaston Berger de Saint-Louis, SENEGAL\\
gane-samb.lo@ugb.edu.sn, ganesamblo@ganesamblo.net}
\address{ LERSTAD, Universit\'e Gaston Berger de Saint-Louis, SENEGAL}

\begin{abstract}
The functional empirical process is a very powerful tool for deriving
asymptotic laws for almost any kind of statistics whenever we know how to
express them into functions of the sample. Since this method seems to be
applied more and more in the very recent future, this paper is intended to provide a
complete but short description and justification of the method and to
illustrate it with a non trivial example using bivariate data. It may also
serve for citation without repeating the arguments.
\end{abstract}

\keywords{functional empirical processes, asymptotic distribution, linear
correlation coefficient}
\subjclass[2000]{Primary 62E20, 62F05. Secondary 62F121, 62GF17}
\maketitle


\section{Introduction}

\label{sec1}

\bigskip \noindent The idea of this paper came out from master and PhD
courses on weak convergence theory of the author in many universities. We
aimed at giving the attendance a powerful tool for finding the asymptotic
normality law of almost any kind of statistics, whenever the latter exists and
the expressions of the statistics in the sample are known. This tool is
the functional empirical process. Usually, this theory is not tauch in
normal courses, but is usually part of dissertation projects. The reason is
the theory is new and it is implicitly assumed that one has to deal with
Donsker classes, Vapnik-Chervonenkis theory (see \cite{vcclass}) or entropy
numbers, in brief that one has to deal with convergence of stochastic
processes. Here, we do not need the big theory in van der vaart and Wellner \cite%
{vaart}, Gaenssler \cite{gaenssler}\ or in Pollard \cite{pollard} or in
books like them.\newline

\bigskip \noindent But at the finite-distribution level, functional
empirical processes can be reasonably easy to use and to justify while
keeping its intrinsic power. We decided to write a paper that explains all
about it at the theoretical level. And next to give a non trivial example.
We choose to deal with the liner correlation that has two interests. On one
hand, we deal with samples of couples of real random variables, to show that
the method goes far beyond real random variables. On another hand, the
computations are indeed somewhat heavy but are also reasonable to allow to
serve as an example for any other case.\newline

\bigskip \noindent It turns now that the paper is not only for master
students but for researchers in Probability and Statistics, in Economics,
and other areas using stochastic methods. As an example, a new theory of
arbitrary Jarque-Berra's type normality law (see \cite{jarque}) is entirely
based on this tool in \cite{gsloJB}.\newline

\bigskip \noindent We are sure that readers of the paper will find it very
useful in their everyday research works.\newline

\bigskip \noindent The remainder of the paper is organized as follows. In
Section \ref{sec2}, we describe the parameter we are estimating, here the
linear correlation coefficient, and its estimator and next establish the
asymptotic normality. In Section \ref{sec3}, we describe our tool, the
empirical process and provide all of its needed features. The proof of the
results will be given finally in Section \ref{sec4}. We end the paper with a
conclusion and recommendations in Section \ref{sec5}.

\section{Weak convergence of the empirical linear correlation coefficient}

\label{sec2}

\bigskip \noindent As said already, we are going to illustrate our tool on
the plug-in estimator of the linear correlation coefficient of \ two random
variable $(X,Y)$ ,with neither of $X$ and $Y$ is degenerated, defined as
follows%
\begin{equation*}
\rho =\frac{\sigma _{xy}}{\sigma _{x}^{2}\sigma _{y}^{2}}
\end{equation*}%
where%
\begin{equation*}
\mu _{x}=\int x\text{ }dP_{X}(x),\text{ }\mu _{y}=\int x\text{ }dP_{X}(x),%
\text{ }\sigma _{xy}=\int (x-\mu _{x})(y-\mu _{y})dP_{(X,Y)}(x,y).
\end{equation*}%
\begin{equation*}
\sigma _{x}^{2}=\int (x-\mu _{x})^{2}dP_{X}(x),\text{ }\sigma _{y}^{2}=\int
(x-\mu _{x})(y-\mu _{y})dP_{X}(y).
\end{equation*}%
We also dismiss the case the case where $\left\vert \rho \right\vert =1$,
for which one of $X$ and $Y$ is an affine function of the other, for example 
$X=aY+b$. \bigskip \noindent It is clear that centering the variables $X$
and $Y$ and normalizing them by their standard deviations $\sigma _{x}$ and $%
\sigma _{y}$ does not change the correlation coefficients $\rho .$ So we may
and do center $X$ and $Y$ at their expectations and normalize them so that
we can and do assume that%
\begin{equation*}
\mu _{x}=\text{ }\mu _{y}=0,\text{ }\sigma _{x}=\sigma _{y}=1.
\end{equation*}

\bigskip \noindent However, we will let these coefficient appear with their
names and we only use their particular values at the conclusion stage.%
\newline

\bigskip \noindent Let us consider the plug-in estimator of $\rho $. To this
end, let $(X_{1},Y_{1}),$ $(X_{2},Y_{2}),...$ be a sequence independent
observations of $(X,Y).$ For each $n\geq 1,$ the plug-in estimator is the
following

\begin{equation*}
\rho _{n}=\left\{ \frac{1}{n}\sum_{i=1}^{n}(X_{i}-\overline{X})(Y_{i}-%
\overline{Y})\right\} \left\{ \frac{1}{n^{2}}\sum_{i=1}^{n}(X_{i}-\overline{X%
})^{2}\times \sum_{i=1}^{n}(X_{i}-\overline{X})^{2}\right\} ^{-1/2}
\end{equation*}

\bigskip \noindent We are going to give the asymptotic theory of $\rho _{n}$
as an estimator of $\rho .$ Introduce the notation

\begin{equation*}
\mu _{(p,x),(q,y)}=E((X-\mu _{x})^{p}(Y-\mu _{y})^{q}),\mu _{4,x}=E(X-\mu
_{x})^{4}\text{, }\mu _{4,x}=E(X-\mu _{x})^{4})
\end{equation*}

\bigskip \noindent Here is our main Theorem

\begin{theorem}
\label{theo1}

\bigskip \noindent Suppose that neither of $X$ and $Y$ is degenerated and
both have finite fourth moments and that $X^{3}Y$ and $XY^{3}$ have finite
expectations. Then, as $n\rightarrow \infty ,$%
\begin{equation*}
\sqrt{n}(\rho _{n}-\rho )\rightsquigarrow N(0,\sigma ^{2}),
\end{equation*}

\bigskip \noindent where 
\begin{eqnarray*}
\sigma ^{2} &=&\sigma _{x}^{-2}\sigma _{y}^{-2}(1+\rho ^{2}/2)\mu
_{(2,x),(2,y)}+\rho ^{2}(\sigma _{x}^{-4}\mu _{4,x}+\sigma _{y}^{-4}\mu
_{4,y})/4 \\
&&-\rho (\sigma _{x}^{-3}\sigma _{y}^{-1}\mu _{(3,x),(1,y)}+\sigma
_{x}^{-1}\sigma _{y}^{-3}\mu _{(1,x),(3,y)})
\end{eqnarray*}
\end{theorem}

\bigskip \noindent This result enables to test independence between $X$ and $%
Y$, or to test non linear correlation in the following sense.

\bigskip

\begin{theorem}
\label{theo2} Suppose that the assumptions of Theorem \ref{theo1} hold. Then%
\newline
\bigskip \noindent \textbf{(1)} If $X$ and $Y$ are not linearly correlated,
that is $\rho =0,$we have 
\begin{equation*}
\sqrt{n}\rho _{n}\rightsquigarrow N(0,\sigma _{1}^{2}),
\end{equation*}

\bigskip \noindent where 
\begin{equation*}
\sigma _{1}^{2}=\sigma _{x}^{-2}\sigma _{y}^{-2}\mu _{(2,x),(2,y)}.
\end{equation*}

\bigskip \noindent \textbf{(2)} If $X$ and $Y$ are independent, then $\rho
=0,$ and\bigskip 
\begin{equation*}
\sqrt{n}\rho _{n}\rightsquigarrow N(0,1)
\end{equation*}
\end{theorem}

\section{The Functional Empirical Process and Other Tools}

\label{sec3}

\bigskip \noindent Let $Z_{1}$, $Z_{2}$, ... be a sequence of independent
copies of a random variable $Z$ defined on the same probability space with
values on some metric space $(S,d)$. Define for each $n\geq 1,$ the
functional empirical process by 
\begin{equation*}
\mathbb{G}_{n}(f)=\frac{1}{\sqrt{n}}\sum_{i=1}^{n}(f(Z_{i})-\mathbb{E}%
f(Z_{i})),
\end{equation*}

\bigskip \noindent where $f$ is a real and measurable function defined on $%
\mathbb{R}$ such that

\begin{equation}
\mathbb{V}_{Z}(f)=\int \left( f(x)-\mathbb{P}_{Z}(f)\right)
^{2}dP_{Z}(x)<\infty ,  \label{var}
\end{equation}%
which entails

\begin{equation}
\mathbb{P}_{Z}(\left\vert f\right\vert )=\int \left\vert f(x)\right\vert
dP_{Z}(x)<\infty \text{.}  \label{esp}
\end{equation}

\bigskip \noindent Denote by $\mathcal{F}(S)$ - $\mathcal{F}$ for short -
the class of real-valued measurable functions that are defined on S such
that (\ref{var}) holds. The space $\mathcal{F}$ , when endowed with the
addition and the external multiplication by real scalars, is a linear space.
Next, it remarkabke that $\mathbb{G}_{n}$ is linear on $\mathcal{F}$, that
is for $f$ and $g$ in $\mathcal{F}$ and for $(a,b)\in \mathbb{R}{^{2}}$, we
have

\begin{equation*}
a\mathbb{G}_{n}(f)+b\mathbb{G}_{n}(g)=\mathbb{G}_{n}(af+bg).
\end{equation*}

\bigskip \noindent We have this result

\begin{lemma} \label{lemma.tool.1}
\bigskip Given the notation above, then for any finite number of elements $%
f_{1},...,f_{k}$ of $\mathcal{S},k\geq 1,$ we have

\begin{equation*}
^{t}(\mathbb{G}_{n}(f_{1}),...,\mathbb{G}_{n}(f_{k}))\rightsquigarrow 
\mathcal{N}_{k}(0,\Gamma (f_{i},f_{j})_{1\leq i,j\leq k}),
\end{equation*}

\bigskip \noindent where 
\begin{equation*}
\Gamma (f_{i},f_{j})=\int \left( f_{i}-\mathbb{P}_{Z}(f_{i})\right) \left(
f_{j}-\mathbb{P}_{Z}(f_{j})\right) d\mathbb{P}_{Z}(x),1\leq ,j\leq k.
\end{equation*}
\end{lemma}

\bigskip \noindent \textbf{Proof}. It is enough to use the Cram\'{e}r-Wold
Criterion (see for example \cite{billinsgley68}, page 45), that
is to show that for any $a=^{t}(a_{1},...,a_{k})\in \mathbb{R}^{k},$ by
denoting $T_{n}=^{t}(\mathbb{G}_{n}(f_{1}),...,\mathbb{G}_{n}(f_{k})),$ we
have $<a,T_{n}>\rightsquigarrow <a,T>$ where $T$ follows the $\mathcal{N}%
_{k}(0,\Gamma (f_{i},f_{j})_{1\leq i,j\leq k})$\ law and $<\circ ,\circ >$
stands for the usual product scalar in $\mathbb{R}^{k}.$ But, by the
standard central limit theorem in $\mathbb{R}$, we have%
\begin{equation*}
<a,T_{n}>=\mathbb{G}_{n}\left( \sum\limits_{i=1}^{k}a_{i}f_{i}\right)
\rightsquigarrow N(0,\sigma _{\infty }^{2}),
\end{equation*}

\bigskip \noindent where, for $g=\sum_{1\leq i\leq k}a_{i}f_{i},$%
\begin{equation*}
\sigma _{\infty }^{2}=\int \left( g(x)-\mathbb{P}_{Z}(g)\right) ^{2}dP_{Z}(x)
\end{equation*}

\bigskip \noindent and this easily gives%
\begin{equation*}
\sigma _{\infty }^{2}=\sum\limits_{1\leq i,j\leq k}a_{i}a_{j}\Gamma
(f_{i},f_{j}),
\end{equation*}%
so that $N(0,\sigma _{\infty }^{2})$ is the law of $<a,T>.$ The proof is
finished.

\subsection{How to use the tool}

\bigskip \noindent We usually work with usual asymptotic statistics on $%
\mathbb{R}^{k}.$ Once we have our sample $Z_{1},Z_{2},...$ as random
variables defined in the same probability space with values in $\mathbb{R}%
^{k},$ the studied statistics, say $T_{n},$ is usually a combinations of
expressions of the form%
\begin{equation*}
H_{n}=\frac{1}{n}\sum\limits_{i=1}^{k}H(Z_{i})
\end{equation*}

\bigskip \noindent for $H\in \mathcal{F}.$ We use this simple expansion, for 
$\mu (H)=\mathbb{E}H(Z),$ 
\begin{equation}
H_{n}=\mu (H)+n^{-1/2}\mathbb{G}_{n}(H).  \label{expan}
\end{equation}

\bigskip \noindent We have that $\mathbb{G}_{n}(H)$ is asymptotically
bounded in probability since $\mathbb{G}_{n}(H)$ weakly converges to, say $%
M(H)$ and then by the continuous mapping theorem $\left\Vert \mathbb{G}%
_{n}(H)\right\Vert \rightsquigarrow \left\Vert M(H)\right\Vert .$ Since all
the $\mathbb{G}_{n}(H)$ are defined on the same probability space, we get
for all $\lambda >0,$ by the assertion of the Portmanteau Theorem for
concerning open sets,%
\begin{equation*}
\lim \sup_{n\rightarrow \infty }P(\left\Vert \mathbb{G}_{n}(H)\right\Vert
>\lambda )\leq P(\left\Vert M(H)\right\Vert >\lambda )
\end{equation*}

\bigskip \noindent and then 
\begin{equation*}
\lim \inf_{\lambda \rightarrow \infty }\lim \sup_{n\rightarrow \infty
}P(\left\Vert \mathbb{G}_{n}(H)\right\Vert >\lambda )\leq \lim \sup
P(\left\Vert M(H)\right\Vert >\lambda )=0.
\end{equation*}

\bigskip \noindent From this, we use the big $O_{\mathbb{P}}$ notation, that
is $\mathbb{G}_{n}(H)=O_{\mathbb{P}}(1).$ Formula (\ref{expan}) becomes%
\begin{equation*}
H_{n}=\mu (H)+n^{-1/2}\mathbb{G}_{n}(H)=\mu (H)+O_{\mathbb{P}}(n^{-1/2})
\end{equation*}

\bigskip \noindent and we will be able to use the delta method. Indeed, let $%
g:\mathbb{R}\longmapsto \mathbb{R}$ be continuously differentiable on a neighborhood of $\mu (H).$
The mean value theorem leads to%
\begin{equation}
g(H_{n})=g(\mu (H))+g^{\prime }(\mu _{n}(H))\text{ }n^{-1/2}\mathbb{G}_{n}(H)
\label{expan01}
\end{equation}

\bigskip \noindent where 
\begin{equation*}
\mu _{n}(H)\in \lbrack (\mu (H)+n^{-1/2}\mathbb{G}_{n}(H))\wedge \mu
(H),(\mu (H)+n^{-1/2}\mathbb{G}_{n}(H))\vee \mu (H)]
\end{equation*}

\bigskip \noindent so that 
\begin{equation*}
\left\vert \mu _{n}(H)-\mu (H)\right\vert \leq n^{-1/2}\mathbb{G}_{n}(H)=O_{%
\mathbb{P}}(n^{-1/2}).
\end{equation*}

\bigskip \noindent Then $\mu _{n}(H)$ converges to $\mu _{n}(H)$ in
probability (denoted $\mu _{n}(H)$ $\rightarrow _{\mathbb{P}}\mu (H)).$ But
the convergence in probability to a constant is equivalent to the weak
convergence. Then $\mu _{n}(H)$ $\rightsquigarrow \mu (H).$ Using again the
continuous mapping theorem, $g^{\prime }(\mu _{n}(H))\rightsquigarrow
g^{\prime }(\mu (H))$ which in tern yields $g^{\prime }(\mu
_{n}(H))\rightarrow _{\mathbb{P}}g^{\prime }(\mu (H))$ by the
characterization of the weak converngence to a constant. Now (\ref{expan01})
becomes%
\begin{eqnarray*}
g(H_{n}) &=&g(\mu (H))+(g^{\prime }(\mu (H)+o_{P}(1))\text{ }n^{-1/2}\mathbb{%
G}_{n}(H) \\
&=&g(\mu (H))+g^{\prime }(\mu (H)\times \text{ }n^{-1/2}\mathbb{G}%
_{n}(H)+o_{P}(1))\text{ }n^{-1/2}\mathbb{G}_{n}(H) \\
&=&g(\mu (H))+\text{ }n^{-1/2}\mathbb{G}_{n}(g^{\prime }(\mu
(H)H)+o_{P}(n^{-1/2})
\end{eqnarray*}

\bigskip \noindent We arrive at the final expansion%
\begin{equation}
g(H_{n})=g(\mu (H))+\text{ }n^{-1/2}\mathbb{G}_{n}(g^{\prime }(\mu
(H)H)+o_{P}(n^{-1/2}).  \label{expanFinal}
\end{equation}

\bigskip \noindent The method consists in using the expansion (\ref%
{expanFinal}) as many times as needed and next to do some algebra on these
expansions. By using the same techniques as above, we have these three
formulas

\begin{lemma} \label{lemma.tool.2}
Let ($A_{n})$ and ($B_{n})$ be two sequences of real valued random variables
defined on the same probability space holding the sequence $Z_{1}$, $Z_{2}$,
.. Let A and B be two real numbers and let $L(z)$ and $H(z)$ be two
real-valued functions$\ of$ $z\in S.$ Suppose that $A_{n}=A+n^{-1/2}\mathbb{G}_{n}(L)+o_{\mathbb{P}}(n^{-1/2})$ and $B_{n}=B+n^{-1/2}\mathbb{G}%
_{n}(H)+o_{\mathbb{P}}(n^{-1/2}).$ Then

\begin{equation*}
A_{n}+B_{n}=A+B+n^{-1/2}\mathbb{G}_{n}(L+H)+o_{\mathbb{P}}(n^{-1/2}),
\end{equation*}

\bigskip \noindent

\begin{equation*}
A_{n}B_{n}=AB+n^{-1/2}\mathbb{G}_{n}(BL+AH) + o_{\mathbb{P}}(n^{-1/2}).
\end{equation*}

\bigskip \noindent and if $B\neq 0$,

\begin{equation*}
\frac{A_{n}}{B_{n}}=\frac{A}{B}+n^{-1/2}\mathbb{G}_{n}(\frac{1}{B}L-\frac{A}{%
B^{2}}H)+ o_{\mathbb{P}}(n^{-1/2}).
\end{equation*}
\end{lemma}

\bigskip \noindent By putting together all the described steps in a smart
way, the methodology will lead us to a final result of the form%
\begin{equation*}
T_{n}=t+n^{-1/2}\mathbb{G}_{n}(h)+o_{P}(n^{-1/2})
\end{equation*}%
which entails the weak convergence%
\begin{equation*}
\sqrt{n}(T_{n}-t)=\mathbb{G}_{n}(h)+o_{P}(1)\rightsquigarrow N(0,\Gamma
(h,h)).
\end{equation*}

\bigskip \noindent We are now in position to apply right here the
methodology on the empirical linear correlation coefficient.

\section{Proofs of the results}

\label{sec4}

\bigskip \noindent We are going to use the function empirical process based
on the observations $(X_{i},Y_{i}),i=1,2,...$ that are independent copies of 
$(X,Y)$. Write%
\begin{equation*}
\rho _{n}^{2}=\frac{\frac{1}{n}\sum_{i=1}^{n}X_{i}Y_{i}-\overline{X}\text{ }%
\overline{Y}}{\left\{ \frac{1}{n}\sum_{i=1}^{n}X_{i}^{2}-\overline{X}%
^{2}\right\} ^{1/2}\left\{ \frac{1}{n}\sum_{i=1}^{n}Y_{i}^{2}-\overline{Y}%
^{2}\right\} ^{1/2}}=\frac{A_{n}}{B_{n}}.
\end{equation*}

\bigskip \noindent Let us say for once that all the functions of $(X,Y)$
that will appear below are measurable and have finite second moments. Let us
handle separately the numerator and denominator. To treat $A_{n},$ using the
empirical process implies that

\begin{equation}
\left\{ 
\begin{tabular}{l}
$\frac{1}{n}\sum_{i=1}^{n}X_{i}Y_{i}=\mu _{xy}+n^{-1/2}G_{n}(p),$ \\ 
$\overline{X}=\mu _{x}+n^{-1/2}G_{n}(\pi _{1}),$ \\ 
$\overline{Y}=\mu _{y}+n^{-1/2}G_{n}(\pi _{2}),$%
\end{tabular}%
\right.  \label{casAN}
\end{equation}

\bigskip \noindent where $p(x,y)=xy$, $\pi _{1}(x,y)=x,$ $\pi _{2}(x,y)=y$.
From there we use the fact that $G_{n}(g)=O_{P}(1)$ for $\mathbb{E}%
(g(X,Y)^{2})<+\infty $ and get%
\begin{equation}
A_{n}=\mu _{xy}+n^{-1/2}G_{n}(p)-(\mu _{x}+n^{-1/2}G_{n}(\pi _{1}))(\mu
_{y}+n^{-1/2}G_{n}(\pi _{2})).  \label{haut}
\end{equation}%
\bigskip \noindent This leads to 
\begin{equation*}
A_{n}=\sigma _{xy}+n^{-1/2}G_{n}(H_{1})+o_{P}(n^{-1/2})
\end{equation*}

\bigskip \noindent with 
\begin{equation*}
H_{1}(x,y)=p(x,y)-\mu _{x}\pi _{2}-\mu _{y}\pi _{1.}
\end{equation*}

\bigskip \noindent Next, we have to handle $B_{n}.$ Since the roles of $%
\left\{ \frac{1}{n}\sum_{i=1}^{n}X_{i}^{2}-\overline{X}^{2}\right\} ^{1/2}$
and of $\left\{ \frac{1}{n}\sum_{i=1}^{n}Y_{i}^{2}-\overline{Y}^{2}\right\}
^{1/2}$ ar symmetrical, we treat one of them and extend the results to the
other. Consider $\left\{ \frac{1}{n}\sum_{i=1}^{n}X_{i}^{2}-\overline{X}%
^{2}\right\} ^{1/2}.$ From (\ref{casAN}), use the delta method to get%
\begin{equation*}
\overline{X}^{2}=\left( \mu _{x}+n^{-1/2}G_{n}(\pi _{1})\right) ^{2}=\mu
_{x}^{2}+2\mu _{x}n^{-1/2}G_{n}(\pi _{1})+o_{P}(n^{-1/2})
\end{equation*}

\bigskip \noindent that is%
\begin{equation*}
\overline{X}^{2}=\left( \mu _{x}+n^{-1/2}G_{n}(\pi _{1})\right) ^{2}=\mu
_{x}^{2}+n^{-1/2}G_{n}(2\mu _{x}\pi _{1})+o_{P}(n^{-1/2}).
\end{equation*}

\bigskip \noindent From there, we get%
\begin{eqnarray*}
\frac{1}{n}\sum_{i=1}^{n}X_{i}^{2}-\overline{X}^{2}
&=&m_{2,x}+n^{-1/2}G_{n}(\pi _{1}^{2})-\overline{X}^{2} \\
&=&m_{2,x}-\mu _{x}^{2}+n^{-1/2}G_{n}(\pi _{1}^{2}-2\mu _{x}\pi
_{1})+o_{P}(n^{-1/2}) \\
&=&\sigma _{x}^{2}+n^{-1/2}G_{n}(\pi _{1}^{2}-2\mu _{x}\pi
_{1})+o_{P}(n^{-1/2}).
\end{eqnarray*}

\bigskip \noindent Using the Delta-method once again leads to 
\begin{equation*}
\left\{ \frac{1}{n}\sum_{i=1}^{n}X_{i}^{2}-\overline{X}^{2}\right\}
^{1/2}=\sigma _{x}+n^{-1/2}G_{n}(\frac{1}{2\sigma _{x}}\left\{ \pi
_{1}^{2}-2\mu _{x}\pi _{1}\right\} )+o_{P}(n^{-1/2}).
\end{equation*}

\bigskip \noindent In a similar way, we get%
\begin{equation*}
\left\{ \frac{1}{n}\sum_{i=1}^{n}Y_{i}^{2}-\overline{Y}^{2}\right\}
^{1/2}=\sigma _{y}+n^{-1/2}G_{n}(\frac{1}{2\sigma _{y}}\left\{ \pi
_{2}^{2}-2\mu _{y}\pi _{2}\right\} )+o_{P}(n^{-1/2}).
\end{equation*}

\bigskip \noindent We arrive at%
\begin{eqnarray*}
B_{n} &=&\left\{ \frac{1}{n}\sum_{i=1}^{n}X_{i}^{2}-\overline{X}^{2}\right\}
^{1/2}\left\{ \frac{1}{n}\sum_{i=1}^{n}Y_{i}^{2}-\overline{Y}^{2}\right\}
^{1/2} \\
&=&\sigma _{x}\sigma _{y}+n^{-1/2}G_{n}(\frac{\sigma _{y}}{2\sigma _{x}}%
\left\{ \pi _{1}^{2}-2\mu _{x}\pi _{1}\right\} +\frac{\sigma _{x}}{2\sigma
_{y}}\left\{ \pi _{2}^{2}-2\mu _{y}\pi _{2}\right\} )+o_{P}(n^{-1/2}).
\end{eqnarray*}

\bigskip \noindent By setting%
\begin{equation*}
H_{2}(x,y)=\frac{\sigma _{y}}{2\sigma _{x}}\left\{ \pi _{1}^{2}-2\mu _{x}\pi
_{1}\right\} +\frac{\sigma _{x}}{2\sigma _{y}}\left\{ \pi _{2}^{2}-2\mu
_{y}\pi _{2}\right\}
\end{equation*}

\bigskip \noindent we have%
\begin{equation}
B_{n}=\sigma _{x}\sigma _{y}+n^{-1/2}G_{n}(H_{2})+n^{-1/2}.  \label{bas}
\end{equation}

\bigskip \noindent Now, combining (\ref{haut}) and (\ref{bas}) and using
Lemma 1 yield%
\begin{equation*}
\sqrt{n}(\rho _{n}^{2}-\rho ^{2})=n^{-1/2}G_{n}(\frac{1}{\sigma _{x}\sigma
_{y}}H_{1}-\frac{\sigma _{xy}}{\sigma _{x}^{2}\sigma _{y}^{2}}%
H_{2})+o_{P}(1).
\end{equation*}

\bigskip \noindent Put%
\begin{equation*}
H=\frac{1}{\sigma _{x}\sigma _{y}}(p(x,y)-\mu _{x}\pi _{2}-\mu _{y}\pi _{1})-%
\frac{\rho }{\sigma _{x}\sigma _{y}}\left\{ \frac{1}{2\sigma _{x}^{2}}%
\left\{ \pi _{1}^{2}-2\mu _{x}\pi _{1}\right\} +\frac{1}{2\sigma _{y}^{2}}%
\left\{ \pi _{2}^{2}-2\mu _{y}\pi _{2}\right\} \right\} .
\end{equation*}

\bigskip \noindent Now we continue with the centered and normalized case to
get%
\begin{equation*}
H(x,y)=p(x,y)-\frac{\rho }{2}(\pi _{1}^{2}+\pi _{2}^{2})
\end{equation*}

\bigskip \noindent and 
\begin{equation*}
H(X,Y)=XY-\frac{\rho }{2}(X^{2}+Y^{2}).
\end{equation*}

\bigskip \noindent Denote%
\begin{equation*}
\mu _{(p,x),(q,y)}=E((X-\mu _{x})^{p}(Y-\mu _{y})^{q}).
\end{equation*}

\bigskip \noindent We have 
\begin{equation*}
EH(X,Y)=\sigma _{xy}-\rho =0
\end{equation*}

\bigskip \noindent and $varH(X,Y)$ is equal to 
\begin{equation*}
\mu _{(2,x),(2,y)}+\rho ^{2}(\mu _{4,x}+\mu _{4,y})/4-\rho (\mu
_{(3,x),(1,y)}+\mu _{(1,x),(3,y)})+\rho ^{2}\mu _{(2,x),(2,y)}/2
\end{equation*}%
and finally $varH(X,Y)=\sigma _{0}^{2}$ with%
\begin{equation*}
\sigma _{0}^{2}=(1+\rho ^{2}/2)\mu _{(2,x),(2,y)}+\rho ^{2}(\mu _{4,x}+\mu
_{4,y})/4-\rho (\mu _{(3,x),(1,y)}+\mu _{(1,x),(3,y)}).
\end{equation*}

\bigskip \noindent This gives the conclusion that for centered and
normalized $X$ and $Y,$%
\begin{equation*}
\sqrt{n}(\rho _{n}-\rho )\rightsquigarrow N(0,\sigma _{0}^{2}).
\end{equation*}

\bigskip \noindent Next, if we use the normalizing coefficients in $\sigma
_{0}$, we get%
\begin{eqnarray*}
\sigma ^{2} &=&\sigma _{x}^{2}\sigma _{y}^{2}(1+\rho ^{2}/2)\mu
_{(2,x),(2,y)}+\rho ^{2}(\sigma _{x}^{4}\mu _{4,x}+\sigma _{y}^{4}\mu
_{4,y})/4 \\
&&-\rho (\sigma _{x}^{3}\sigma _{y}\mu _{(3,x),(1,y)}+\sigma _{x}\sigma
_{y}^{3}\mu _{(1,x),(3,y)})
\end{eqnarray*}

\bigskip \noindent and we conclude in the general case that%
\begin{equation*}
\sqrt{n}(\rho _{n}-\rho )\rightsquigarrow N(0,\sigma ^{2})
\end{equation*}

\bigskip \noindent The proof of Theorem \ref{theo2} follows by easy
computations under the particular conditions of $\rho $ and under
independence.

\section{Conclusion and recommendations}

This paper was intended to lear how to use the functional empirical process
for deriving asymptotic normality laws. We saw that the method is powerful
and sometimes straightforward. But the computations may be huge.
Fortunately, in each case, we might be able to use specific properties of
studied statistics. In the example we treated, we use the property that the
correlation coefficient remains the same under location and scale parameter
changes. This allowed to reduce the computations to centered and normalized
random variables.

\bigskip

This works has sparked in our mind the ideas of a handbook in which as much
as possible applications of this method in any area of Probability theory
and Statistics will be gathered. We will surely go into it in the future

\bigskip

\begin{center}
\bigskip Acknowledgement
\end{center}

We express our thanks to our last master class in the Mathematical
department of University of Sciences, Techniques and Technologies of Bamako,
in June 2016, where we made the decision to write this paper. Thanks to the
attendance ; Ch\'{e}rif Traor\'{e}, Maryam Traor\'{e}, Paul Akounid, Dian Diarra,
Harouna Sangar\'{e}, Soumaila Demebel\'{e}, Mouminou Diallo. We extend our
thanks to the agency AUF which funded the mission there and the department
of mathematics and the USTTB through Dr Yaya Kon\'{e} and Professor Ouat\'{e}ni Diallo.

\label{sec5}

\end{document}